\documentclass[epj]{svjour}
\usepackage{graphics}
\usepackage{epstopdf}
\usepackage{color}
\usepackage{textcomp}
\begin{document}
\title{Extended generalized recurrence plot quantification of complex circular patterns}
\author{Maik Riedl\inst{1} \and Norbert Marwan\inst{1} \and J\"urgen Kurths\inst{1}\inst{2}\inst{3}
}                     
%
%
\institute{Potsdam Institute for Climate Impact Research, P.O. Box 60 12 03, 14412 Potsdam, Germany \and Humboldt Universit\"at zu Berlin, Department of Physics, Newtonstraße 15, 12489 Berlin, Germany \and Nizhny Novgorod State University, 23 Prospekt Gagarina BLDG 2, 603950 Nizhnij Novgorod, Russia}
\date{Received: date / Revised version: date}
%
\abstract{
The generalized recurrence plot is a modern tool for quantification of complex spatial patterns. 
Its application spans the analysis of trabecular bone structures, Turing patterns, turbulent spatial plankton patterns, and fractals. 
Determinism is a central measure in this framework quantifying the level of regularity of spatial structures.
We show by basic examples of fully regular patterns of different symmetries that this measure underestimates the orderliness of circular patterns resulting from rotational symmetries. 
We overcome this crucial problem by checking additional structural elements of the generalized recurrence plot which is demonstrated with the examples.
Furthermore, we show the potential of the extended quantity of determinism applying it to more irregular circular patterns which are generated by the complex Ginzburg-Landau-equation and which can be often observed in real spatially extended dynamical systems. 
So, we are able to reconstruct the main separations of the system\textquotesingle s parameter space analyzing single snapshots of the real part only, in contrast to the use of the original quantity.
This ability of the proposed method promises also an improved description of other systems with complicated spatio-temporal dynamics typically occurring in fluid dynamics, climatology, biology, ecology, social sciences, etc.
\PACS{
      {05.10.-a}{Computational methods in statistical physics and nonlinear dynamics}   \and
      {05.45.-a}{Nonlinear dynamics and chaos} \and
			{02.50.Sk}{Multivariate analysis} \and
			{05.45.Tp}{Time series analysis} \and
			{89.75.Fb}{Structures and organization in complex systems} \and
			{89.75.Kd}{Patterns}
     } 
} 
\maketitle
\section{Introduction}
\label{intro}
A sufficient quantitative description of complex spatial patterns is still an open question which arises in many fields, such as fluid dynamics, climatology, biology, ecology, and social sciences. 
Often we ask for such quantities in order to automatically detect regime shifts in systems by means of their spatio-temporal dynamics assuming a coupling of spatial patterns and rhythm. 
A promising framework solving this problem is the recurrence plot (RP) analysis \cite{Eckmann1987,Marwan2007a} with its library of measures, the recurrence quantification analysis \cite{Marwan2007a} and the recurrence network analysis \cite{Marwan2009,Gao2009,Gao2010}. 
One of the most useful quantity of those is the measure of determinism ($\Delta$) describing the level of regularity.
There are several methodologies to construct a RP from spatial data: 
(1) the spatial RP \cite{Vasconcelos2006,Prado2014}, (2) in particular the generalized RP (GRP) \cite{Marwan2007b}, and (3) its more computationally performant approximation assuming an isotropic spatial structure \cite{Facchini2009a,Facchini2009b,Mocenni2010,Moccenni2011,Facchini2013}.
The spatial RP separately analyzes the recurrence in regular defined subgroups of the spatial data for a spatially resolved view, whereas the GRP gives a global view on the recurrence of spatial patterns avoiding the selection of the subgroups.
These approaches have been already successively applied to the description of complex spatial structures of trabecular bones during bone loss in osteoporosis \cite{Marwan2007b}; Turing patterns \cite{Facchini2009a,Moccenni2011,Facchini2013,Silva2014}; patterns generated by the Belousov–Zhabotinsky reaction, the complex Ginzburg–Landau equation (GLE), or an extension of the Rosenzweig-MacArthur model \cite{Facchini2009a,Moccenni2011,Tang2014}; the chlorophyll distribution in ocean colonies of plankton given by remote sensed ocean colors \cite{Facchini2009a}; and fractals \cite{Facchini2009a} .
Despite this success a crucial open question remains: Comparing the GRP with the classical RP approach, we expect that fully regular spatial structures lead to $\Delta=1$.
However, in fact we find $\Delta$ values smaller than 1 \cite{Marwan2007b}.
In this work we will treat this problem of underestimation and finally provide a solution.
We will show that we have to check additional structures in the GRP which differ from such the regular $\Delta$ is based on (Sec.\ref{method}).
Further we will show that these structures correspond to specific symmetries of the spatial patterns and that their consideration substantially improves the description of them, especially in the case of complex circular patterns, in relation to the original GRP approach (Sec.\ref{simpleEx}).


\section{\label{method} Method}

\subsection{Generalized recurrence plot (GRP)}
Let us consider a $D$-dimensional data field.
The elements of this field are $m$-dimensional vectors $\vec{x}_{\vec{i}}$ which have to be arranged in a regular $D$-dimensional grid where the $D$-tuples $\vec{i}=(i_1,...,i_D)$ determine the location of the vector in the grid.
An example of this general case is a snapshot of a system which is distributed in a 2-dimensional spatial space.
The elements of the resulting 2-dimensional data field represent the state vectors of the system at the different positions in this spatial space at the considered time stamp.
We encode the pairwise similarity of the vectors $\vec{x}_{\vec{i}}$ by constructing the GRP: 
\begin{equation}\label{eq1a}
R_{\vec{i}\vec{j}}=\theta(\epsilon-\|\vec{x}_{\vec{i}}-\vec{x}_{\vec{j}}\|)\end{equation}
where $\|\bullet\|$ denotes the norm of the vector space and $\epsilon$ defines the threshold of the distance where two states are no longer assumed as similar. 
The Heaviside function $\theta(\bullet)$ encodes this condition of similarity and is 1 for distances smaller than $\epsilon$; otherwise it is 0. 
That is, the GRP is a $2D$-dimensional binary matrix where $R_{\vec{i}\vec{j}}=1$ indicates the assumed similarity of two states at the locations $\vec{i}$ and $\vec{j}$ in the data field.
Zero stands for non-similar states \cite{Marwan2007b}.


\subsection{Structures of the GRP}
\begin{figure*}
\centering
\resizebox{0.75\paperwidth}{!}{\includegraphics{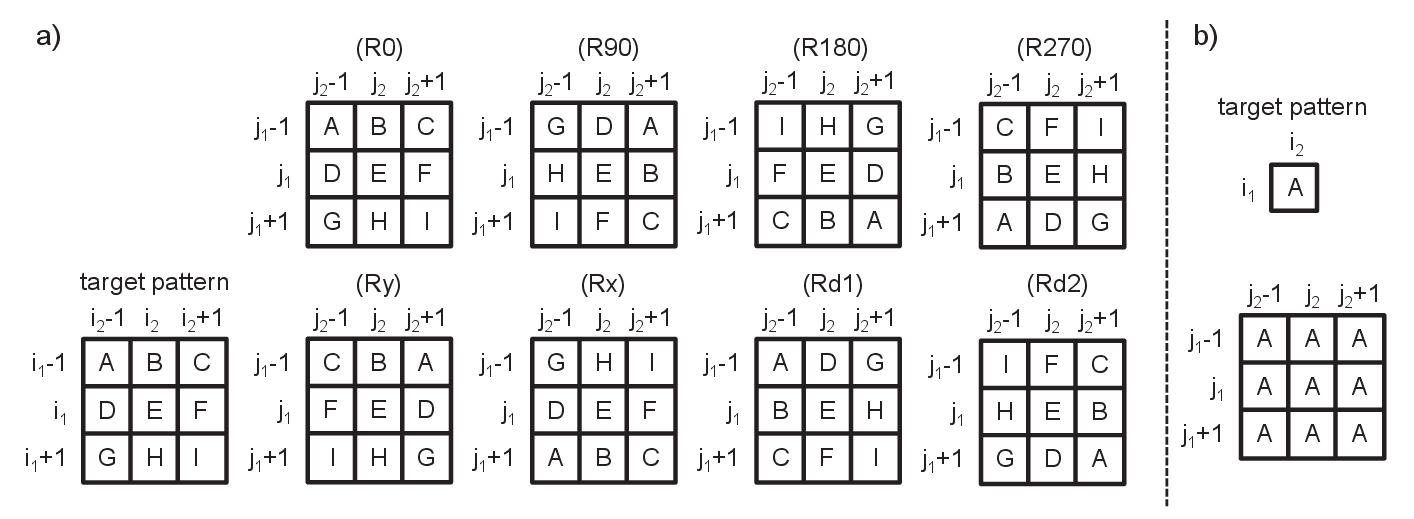}}
\caption{a) Possible cases of recurrence of a target pattern in a 2-dimensional data field where the local neighborhood of the states in the target pattern is preserved: R0 $\equiv$ shift, R90 (R180, R270) $\equiv$ shift + $90^{\circ}$($180^{\circ}$,$270^{\circ}$) rotation, Ry (Rx, Rd1, Rd2) $\equiv$ shift + reflection on the vertical (horizontal, first diagonal, second diagonal) axis. b) Recurrence of a target state in a larger pattern of equal states which  corresponds to a vertical structure of 1 in the GRP (first $D$ coordinates of the representing elements of the GRP are constant).}
\label{fig1}
\end{figure*}

In relation to the classical RP, we consider $D$-dimensional structures in the GRP which are built by neighbored 1. 
The current quantification of the GRP is a straight extension of the classical RP which is mainly based on two classes of such structures, diagonal and vertical structures.
So, $\Delta$ is the proportion of recurrences within the diagonal structures among all recurrences within the GRP.
The diagonal structures have the same orientation as the hyper-surface of identity in the GRP and relates to the occurrence of a $D$-dimensional pattern at two positions of the $D$-dimensional data field with equal orientation.
For $D=2$, this case is illustrated by the target pattern and the pattern R0 in Fig.~\ref{fig1}a, for example.
This can be interpreted as a shift of the pattern from one position of the data field to another one where the mapping preserves the spatial relation of the states in the subsets.
That is, if two elements are neighbors in the reference then the mapped elements are neighbors, too, in the second subset.
But this property is not only given by the shift operation (R0 in Fig.~\ref{fig1}a) but also given by combinations of rotation, reflection and shift as illustrated by the patterns R90, R180, R270, Ry, Rx, Rd1, Rd2 in Fig.~\ref{fig1}a.
These last recurrences are also encoded in the GRP because of the point-wise test of recurrence (Eq.~\ref{eq1a}). 
They also build 2-dimensional structures in the GRP which are different from the diagonal ones.
So, these diagonal-like structures decreases the proportion of 1 in the diagonal structures among all points in the GRP and might be the cause of the aforementioned underestimation of $\Delta$ in the case of regular patterns.
We hypothesize that the additional consideration of these diagonal-like structures will fit this problem.  

Let us demonstrate this for $D=2$ explicitly.
For the simplest formulation, especially in the case of rotations, we consider $D$-dimensional cuboids of different sizes, i.e. squares in this case, which build the more complex shapes of the $D$-dimensional diagonal and diagonal-like structures of neighbored 1 in the GRP.  
These squares are defined by
\begin{equation}\label{eq6}
\theta(B_1+B_2)\prod_{k_1,k_2=b_1}^{b_2}{R_{(\vec{i}+\vec{k},\vec{j}+\vec{k}')}}\equiv 1
\end{equation}
\begin{eqnarray}\label{eq7}
B_1=&&\sum_{s=b_1-1}^{b_2+1}{(1-R_{(\vec{i}+\vec{u},\vec{j}+\vec{u}')})}\nonumber\\
&&\times \sum_{s=b_1-1}^{b_2+1}{(1-R_{(\vec{i}+\vec{v},\vec{j}+\vec{v}')})}
\end{eqnarray}
\begin{equation}\label{eq8}
B_2=\sum_{s=b_1}^{b_2}{(1-R_{(\vec{i}+\vec{w},\vec{j}+\vec{w}')})}\sum_{s=b_1}^{b_2}{(1-R_{(\vec{i}+\vec{z},\vec{j}+\vec{z}')})}
\end{equation}
\begin{eqnarray}\label{eq9}
\vec{u}=(b_1-1,s);&&\vec{v}=(b_2+1,s);\nonumber\\
\vec{w}=(s,b_1-1);&&\vec{z}=(s,b_2+1)
\end{eqnarray}
\begin{equation}\label{eq10}
b_1=\left\{\begin{array}{cc}-l/2+1 & l\mbox{ is even} \\ -l/2+1/2 & l\mbox{ is odd} \end{array}\right.
\end{equation}
\begin{equation}\label{eq11}
b_2=\left\{\begin{array}{cc}l/2 & l\mbox{ is even} \\ l/2-1/2 & l\mbox{ is odd} \end{array}\right.
\end{equation}
The conversion of the vectors $\vec{k}$, $\vec{u}$, $\vec{v}$, $\vec{w}$, and $\vec{z}$ , i.e. $\vec{k}'$, $\vec{u}'$, $\vec{v}'$, $\vec{w}'$ ,and $\vec{z}'$, is given in Tab.~\ref{tab1} in relation to the different mappings.
Compositions of these basic mappings have not to be explicitly considered since they can be represented by the basic mappings.

\begin{table}
\caption{Transformed coordinates in relation to the mapping $M$ of the target pattern in the two dimensional field (cf. Fig.~\ref{fig1}a). If $\vec{k}$, $\vec{u}$, $\vec{v}$, $\vec{w}$, and $\vec{z}$ have the form $\vec{x}=(x_1,x_2)$ then their transformations, $\vec{k}'$, $\vec{u}'$, $\vec{v}'$, $\vec{w}'$, and $\vec{z}'$, are:}
\label{tab1} 
\centering
\begin{tabular}{lll}
\hline\noalign{\smallskip}
Mapping $M$ &\multicolumn{2}{c}{Transformation of the coordinates}\\
&$l$ is odd&$l$ is even\\
\noalign{\smallskip}\hline\noalign{\smallskip}
R0 & $(x_1,x_2)$&$(x_1,x_2)$\\
R90& $(x_2,-x_1)$&$(x_2,-x_1+1)$\\
R270& $(-x_2,x_1)$&$(-x_2+1,x_1)$\\
R180& $(-x_1,-x_2)$&$(-x_1+1,-x_2+1)$\\
Ry & $(x_1,-x_2)$ & $(x_1,-x_2+1)$\\
Rx & $(-x_1,x_2)$ & $(-x_1+1,x_2)$\\
Rd1& $(x_2,x_1)$ & $(x_2,x_1)$\\
Rd2& $(-x_2,-x_1)$ & $(-x_2+1,-x_1+1)$\\
\noalign{\smallskip}\hline
\end{tabular}
\end{table}

The terms $B_1$ and $B_2$ are the boundary conditions of the squared patch in the GRP.
One can see that the original diagonal structure is a special case in this formulation (first line in Tab.~\ref{tab1}). 
In contrast to previous work \cite{Marwan2007b}, the reference point of the quadratic patches, i.e. $\vec{i}$ or $\vec{j}$, is the central point in the case of an uneven size $l$ and otherwise the point in the patch next to the center with the smallest coordinates. 
So the conversions of the coordinates (Tab.~\ref{tab1}) do not depend on the size of the patches which helps to separate the shift given by the vector $(\vec{i},\vec{j})$ in Eq.~\ref{eq6} from the other operations determined by $(\vec{k},\vec{k}')$.
The formulation of the considered structures for arbitrary values of $D$ are given in the appendix (App.~\ref{allD}). 

For the sake of completeness, the second basic structures of the classical recurrence quantification analysis \cite{Marwan2007a}, the vertical structures of 1, are also considered.
They are given by
\begin{equation}\label{eq12}
\theta\left(\sum_{a=1}^D B_a'\right)\prod_{k_1,...,k_D=b_1}^{b_2}{R_{(\vec{i},\vec{j}+\vec{k})}}\equiv 1
\end{equation}
\begin{eqnarray}\label{eq13}
B_a'=&&\sum_{s_1,...,s_{D-1}=b_1-1}^{b_2+1}{(1-R_{(\vec{i},\vec{j}+\vec{u})})}\nonumber\\
&&\times \sum_{s_1,...,s_{D-1}=b_1-1}^{b_2+1}{(1-R_{(\vec{i},\vec{j}+\vec{v})})}
\end{eqnarray}

where $b_1$ and $b_2$ are defined by Eq.~\ref{eq10} and Eq.~\ref{eq11}, respectively.
$\vec{i}$, $\vec{j}$, $\vec{u}$, and $\vec{v}$ are $D$ tuples of coordinates where in the last two ones the $a$-th component is fixed at $b_1-1$ and $b_2+1$, respectively.
The remaining components are the running indexes in Eq.~\ref{eq13}, $s_1$,...,$s_{D-1}$.
For this kind of recurrence structure, the orientation of the pattern plays no role because of the fact that the considered $D$ cuboid consists of equal states.


\subsection{Extended quantification of the recurrence structures}

Quantifying the diagonal like structures of 1 in the GRP, we extend the established measure $\Delta$ in the following way
\begin{equation}\label{eq16}
\Delta_{M}=\frac{\sum_{l=l_{min}}^{N}{l^D P_{M}(l)}}{\sum_{l=1}^{N}{l^D P_{M}(l)}}
\end{equation}
where $M$ indicates the considered mapping (see Tab.~\ref{tab1}).
In this more general formulation, the original measure $\Delta$ is given by $\Delta_{R0}$.
Here, $l^D$ is the volume of the $D$ dimensional cuboid, i.e. the enclosed number of points of the GRP, and $P_{M}(l)$ is the histogram of cuboids' size $l$. 
$N$ is the maximum size of the quadratic patches. 
A sufficient value of $l_{min}$ diminishes artificial contributions which are caused by broad areas of quiet similar values in the data field \cite{Marwan2007a}.
Beside this correction, $l_{min}$ can also use to determine the minimal spatial scale of interest. 
So, $\Delta_M$ is the portion of 1 in the GRP which lie in cuboids of size longer than $l_{min}-1$ which relates to the considered mapping $M$. 
In our analysis of two-dimensional data fields, we set $l_{min}=2$. 

\begin{figure}
\centering
\resizebox{0.35\textwidth}{!}{\includegraphics{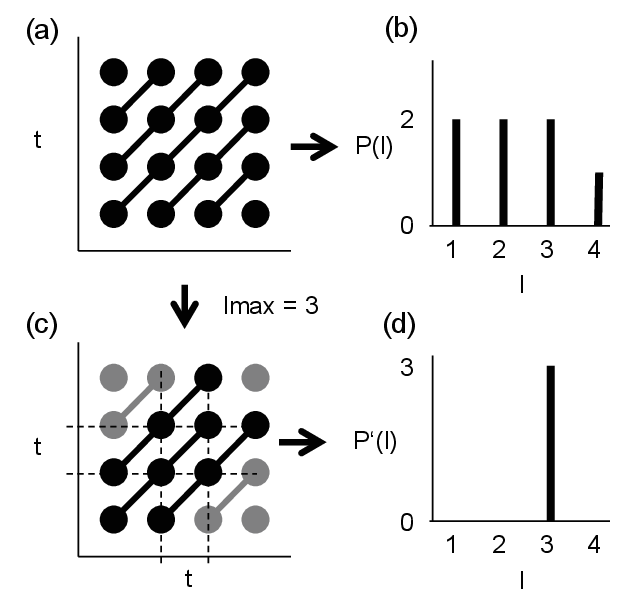}}
\caption{Recurrence plot of a series of four 1 (a) and the corresponding histogram of the line lengths $P(l)$ (b). The definition of the maximum allowed length $l_{max}$ leads to a split of longer lines and the determination of allowed lines characterized by the black color (c). Condition of these lines is that their center is in the interval $[l_{max}/2+1/2,N-l_{max}/2+1/2]$  marked by the dashed lines. Only these lines are used to build the corrected histogram of line length $P'(l)$ (d).}
\label{fig4}
\end{figure}

Calculating $\Delta_M$, a boundary effect has to be respected.
It results from boundary 1 in the GRP which are not able to build a full $D$ dimensional cuboid because of the cutting edges. 
The larger the possible cuboids are the more probably is their cutting. 
This leads to an underestimation of the number of these larger cuboids in the histogram $P_M(l)$ (Eq.~\ref{eq16}) and an overestimation of the number of smaller cuboids which are built from the remaining bounding 1. 
That is, the boundary effect causes an underestimation of $\Delta_M$ (Eq.~\ref{eq16}). 
Therefore, we propose a correction which diminishes this boundary effect and guarantees consistent values of $\Delta_M$, e.g. $\Delta_{R0}=1$ in the case of always recurrence. 
The idea of this correction is the limitation of the maximal size $l_{max}$ of the considered cuboids splitting larger structures of 1 in the GRP which are at risk of cutting. Further, this $l_{max}$ is used to define the range of the border of 1 in the GRP which are expected to be remains of cutting and therefore are not rated as center of countable cuboids. 
So, the underestimation of the larger cuboids and the overestimation of the smaller cuboids is diminished in the construction of $P_M(l)$ (Eq.~\ref{eq16}). 
$l_{max}$ has to be set between $l_{min}$ (Eq.~\ref{eq16}) and N which defines the indexes of the bordering 1, $i<l_{max}/2+1/2$ or $i>N-l_{max}/2+1/2$. 
An illustration of this correction is given for the example of the one-dimensional case (Fig.~\ref{fig4}). 
Here, a 4$\times$4 recurrence matrix with only 1 is displayed by black dots (Fig.~\ref{fig4}a). 
For $D=1$, the cuboids are lines. 
Counting the diagonal lines of specific length marked by the black solid lines, we get $P_{R0}(l)$ of Eq.~\ref{eq16} (Fig.~\ref{fig4}b). 
Although only lines of length 4 are probably, the cutting leads to diagonal lines of length 1, 2, and 3. 
If we set $l_{max}$ to 3 then the larger diagonal, the main diagonal, is split into two parts of length 3 and 1. 
Further, the range of the center of the countable lines is defined and only contains the inner points (c.f. Fig.~\ref{fig4}c). 
Now, the resulting histogram $P_{R0}(l)$ only shows one line and the corresponding value of $\Delta$ is 1 (for $l_{min}=2$) as we expect for always recurrence.


\section{\label{simpleEx} Application to different spatial patterns}

\subsection{Regular patterns and noise}

\begin{figure}
\centering
\resizebox{0.5\textwidth}{!}{\includegraphics{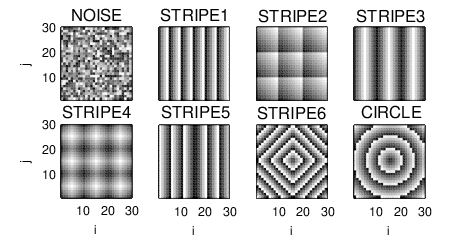}}
\caption{Examples of 2-dimensional white noise (NOISE) and regular spatial patterns (App.~\ref{regEx}): with asymmetric shape of the states in one direction (STRIPE1) and two  directions (STRIPE2), with symmetric shape in one direction (STRIPE3) and two directions (STRIPE4), with asymmetric shape in one direction and a reflection along the vertical median line (STRIPE5), with asymmetric shape in two directions and reflections along the vertical and horizontal median line (STRIPE6), with concentric circles around the middle (CIRCLE).}
\label{fig3} 
\end{figure}

First, let us apply this extended quantification of the GRP and the Laminarity ($\Lambda$), a further important measure of the RP basing on vertical line structures (App.~\ref{LAM}), to regular 2-dimensional spatial patterns as well as noise (Fig.~\ref{fig3}, App.~\ref{regEx}) where $D=2$ and $m=1$.
The size of the squared data field is set to $N=30$.
For each STRIPE pattern, $\epsilon$ (Eq.~\ref{eq1a}) is set to the smallest non-zero distance between the states resulting in recurrence rates ($h$) which are shown in Tab.\ref{tab2}.
The use of a fixed threshold $\epsilon$ guarantees that all pairs of equal states are valued, whereas the determination of $\epsilon$ by means of a fixed $h$ can leave out of consideration parts of this pairs.
In the case of only equal values in the data field (CONSTANT), this error is maximal, since the fixed value of $h$ would always be smaller than the expected value of 1.
The pattern CIRCLE and NOISE do not show this high amount of equal values as the STRIPE patterns.
Therefore, we choose here a greater value of $\epsilon$ than the minimal non-zero distance in order to get values of $h$ which are in the range of the STRIPE pattern (cf. Tab.~\ref{tab2}).
Correcting the boundary effect of $\Delta_M$, the maximal size is set to $l_{max}=10$ grid points.

\begin{table*}
\caption{Quantities of the GRP for the different pattern (cf. Fig.~\ref{fig3}): $\Delta_{M}$ - determinism which relates to the specific mapping $M$ (see Tab.~\ref{tab1} and Eq.~\ref{eq16}); $h$ - recurrence rate; $\Lambda$ - laminiarity (Eq.~\ref{eq17}); $\Delta_{\Sigma M}$,  - cumulative detection of the determinism considering all mappings ($\Delta_{all}$), only rotations ($M=$R0, R90, R180, and R270; $\Delta_{rot}$), and only reflections ($M=$Ry, Rx, Rd1, Rd2; $\Delta_{ref}$)}
\label{tab2}
\centering       
\begin{tabular}{l|llllllll|lll|ll}
\hline\noalign{\smallskip}
Pattern & \multicolumn{8}{c}{$\Delta_{M}$}         &\multicolumn{3}{c}{$\Delta_{\Sigma M}$}&$h$&$\Lambda$\\
        &R0  &R90 &R180&R270&Ry  &Rx  &Rd1 &Rd2 &all &rot &ref &    &    \\
\noalign{\smallskip}\hline\noalign{\smallskip}
CONSTANT&1   &1   &1   &1   &1   &1   &1   &1   &1   &1   &1   &1   &1   \\
NOISE   &0.06&0   &0.01&0   &0.01&0   &0.01&0.02&0.15&0.1 &0.09&0.1 &0   \\
STRIPE1 &1   &0   &0   &0   &0   &1   &0   &0   &1   &1   &1   &0.2 &0   \\
STRIPE2 &0.99&0   &0   &0   &0   &0   &1   &0   &1   &1   &1   &0.07&0   \\
STRIPE3 &0.82&0   &0.75&0   &0.75&0.82&0   &0   &1   &1   &1   &0.18&0   \\
STRIPE4 &0.56&0.51&0.47&0.51&0.51&0.51&0.56&0.47&0.98&0.98&0.98&0.13&0   \\
STRIPE5 &0.73&0.05&0.73&0.05&0.73&0.73&0.05&0.05&1   &1   &1   &0.2 &0.07\\
STRIPE6 &0.35&0.35&0.35&0.35&0.35&0.35&0.35&0.35&0.99&0.99&0.99&0.2 &0   \\
CIRCLE &0.23&0.23&0.23&0.23&0.23&0.23&0.23&0.23&0.94&0.91&0.91&0.19&0   \\
\noalign{\smallskip}\hline
\end{tabular}
\end{table*}

The results are given in Tab.~\ref{tab2}.
First of all, the first column shows the aforementioned problem of the original quantification of the GRP by means of $\Delta$ ($\Delta\equiv \Delta_{R0}$).
Despite the correction of the boundary effect, the values of $\Delta_{R0}$ are clearly smaller than 1 as for STRIPE3 to CIRCLE although the value one is expected for these regular patterns. 
Except the constant pattern (different $\Lambda$, Tab.~\ref{tab2}) and the noise (no symmetries), an increasing number of symmetries reduces the value of $\Delta_{R0}$, but bears non-zero values of other components of $\Delta$, e.g. $\Delta_{R180}=0.75$ for STRIPE3. 
Unfortunately, these non-zero components cannot be added in order to get the expected value of 1 for the overall determinism, due to a strong overlapping of the different 
structures of 1 in the GRP.
So the recurrence structures, related to $\Delta_{R0}$ and $\Delta_{Rx}$, share the same 1 in the GRP in the case of STRIPE1, for example.   
In particular, it seems, that each rotational mapping (R0, R90, R180, and R270) has a corresponding operation in the group of reflections (Rx, Ry, Rd1, and Rd2) resulting in the same values of $\Delta_M$ (cf. Tab.~\ref{tab2}).
In order to overcome this overlapping, $\Delta_{\Sigma M}$ is determined by a cumulative quantification where the single $\Delta_M$ are successively calculated using the remaining recurrence points which are not part of a valued cuboid in the GRP by the previous step.
This cumulative detection $\Delta_{\Sigma M}$ leads to values which are equal or close to the expected 1 in the case of regular patterns (Tab.~\ref{tab2}).
In particular, the consideration of all mappings $\Delta_{all}$ do not lead to an improvement in comparison to the only use of the rotational mappings $\Delta_{rot}$ as well as the group of reflections $\Delta_{ref}$.
This finding relates to the aforementioned observation that there are corresponding operations in the two groups.
Beside this, the results also show the equivalence of $\Delta_{R90}$ and $\Delta_{R270}$ (Tab.~\ref{tab2}), since the mapping $R270$ is the inverse of $R90$ if we consider not only a clockwise rotation but also an anti-clockwise one.
The case of noise shows that the selected value of $l_{min}=2$ is enough to diminish random effects for $D=2$.
Finally, the values of $\Delta_M$ and $\Delta_{\Sigma M}$ are similar in almost all cases for a fixed recurrence rate of 0.2.
Only the case STRIPE2 leads to a high laminarity resulting in additional non-zero parts of $\Delta_M$.


\subsection{\label{complexEx} Complex GLE}

\begin{figure}
\centering
\resizebox{0.5\textwidth}{!}{\includegraphics{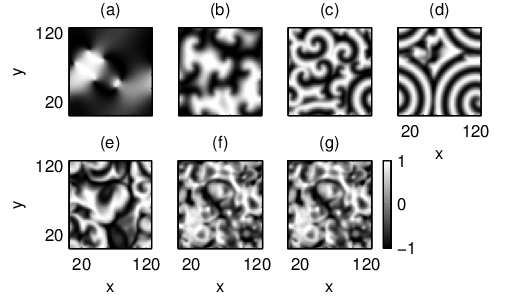}}
\caption{Real part of the $1500$-th time step of simulations of the complex Ginzburg-Landau equation (Eq.~\ref{eq18}). The parameters (c,b) are: a) $(0,-0.25)$, b) $(-0.3,-2.75)$, c) $(0.3,-1.25)$, d) $(0.4,-2.5)$, e) $(0.7,-2.25)$, f) $(1.2,-2.75)$, and g) $(1.9,-0.75)$}
\label{fig5} 
\end{figure}

In the previous section, we tested the proposed extension of the GRP quantification on regular spatial patterns and random ones.
But in nature we find a whole spectrum of spatial patterns which are between these extrema.
A prominent theoretical generator of such complex patterns is the complex GLE which is one of the most-studied nonlinear systems in physics generalizing a variety of phenomena in spatially extended systems from spiral waves to turbulence \cite{Aranson2002}. 
We want to demonstrate the ability of the proposed method by a differentiation of the complex spatial patterns resulting from the dynamics of the complex GLE in order to get information on the observed system.
The system is defined by:
\begin{eqnarray}\label{eq18}
\partial_t A(x,y,t)=&&A(x,y,t)+(1+ib)\Delta A(x,y,t)\nonumber\\
&&-(1+ic)\vert A(x,y,t)\vert ^2 A(x,y,t)
\end{eqnarray}
where $A$ is a field of complex numbers.  
The parameters $b$ and $c$ characterize linear and nonlinear dispersion, respectively.
Simulating this system, we use an exponential time-differencing fourth-order Runge-Kutta scheme \cite{Kassam2003} with time steps of size $1/8$.  
The size of the squared field is $128$ elements which equidistantly sample the spatial range of $[0,100] \times [0,100]$.
We consider periodic boundary conditions and the initial condition is set to white noise with standard deviation of $0.1$.
Typical examples of resulting spatial patterns are shown in Fig.~\ref{fig5}.

\begin{figure*}
\centering
\resizebox{0.75\paperwidth}{!}{\includegraphics{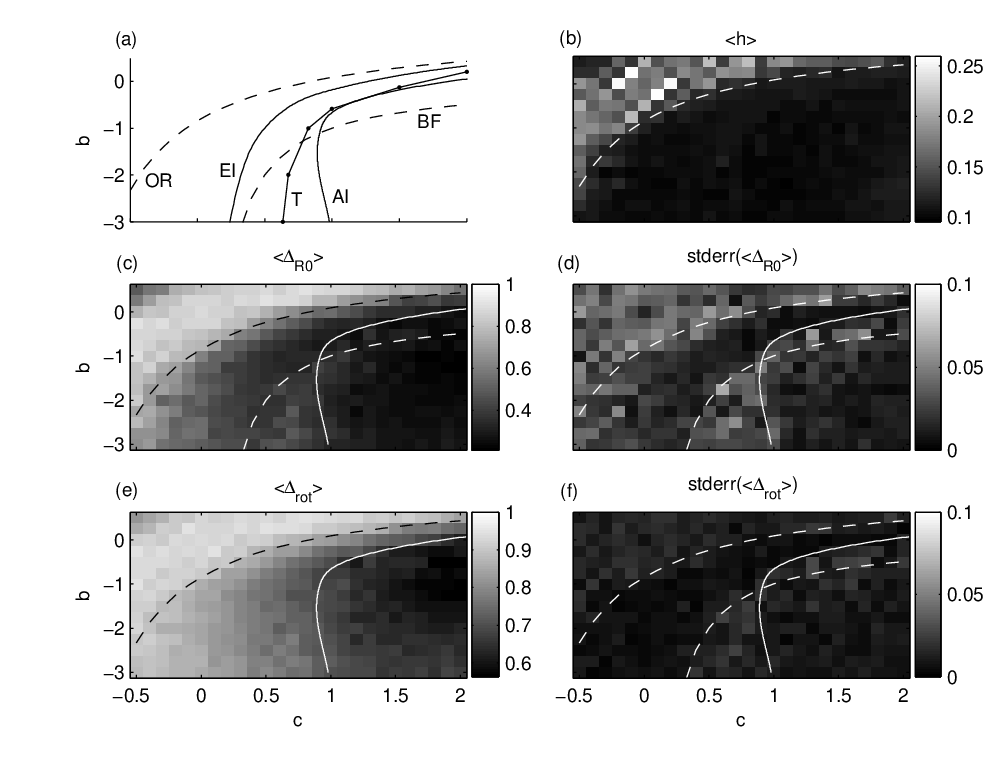}}
\caption{(a) Scheme of transition lines of the complex GLE which are copied from Aranson \& Kramer~\cite{Aranson2002} with: the transition to oscillatory regimes (OR), i.e. bounded states, from the top to the bottom; the Eckhaus instability (EI) and the absolute instability (AI) enclosing the condition for spiral structures; T the transition from frozen states to turbulence from the left to the right; and the Benjamin-Feir-Newell line (BF) dividing phase turbulence above and defect turbulence below, (b) mean recurrence rate $h$ for each parameter combination, (c) mean of the boundary corrected original measure of determinism $\Delta_{R0}$, (d) standard error of the values in c, (e) mean of the boundary corrected extended measure of determinism $\Delta_{rot}$, (f) standard error of the values in e. The lines in b, c, d, e, and f are adapted from panel a and characterize dominant gradients in the gray scale plots. The different colors of the line are only used for a higher contrast.}
\label{fig6} 
\end{figure*}

\begin{figure*}
\centering
\resizebox{0.75\paperwidth}{!}{\includegraphics{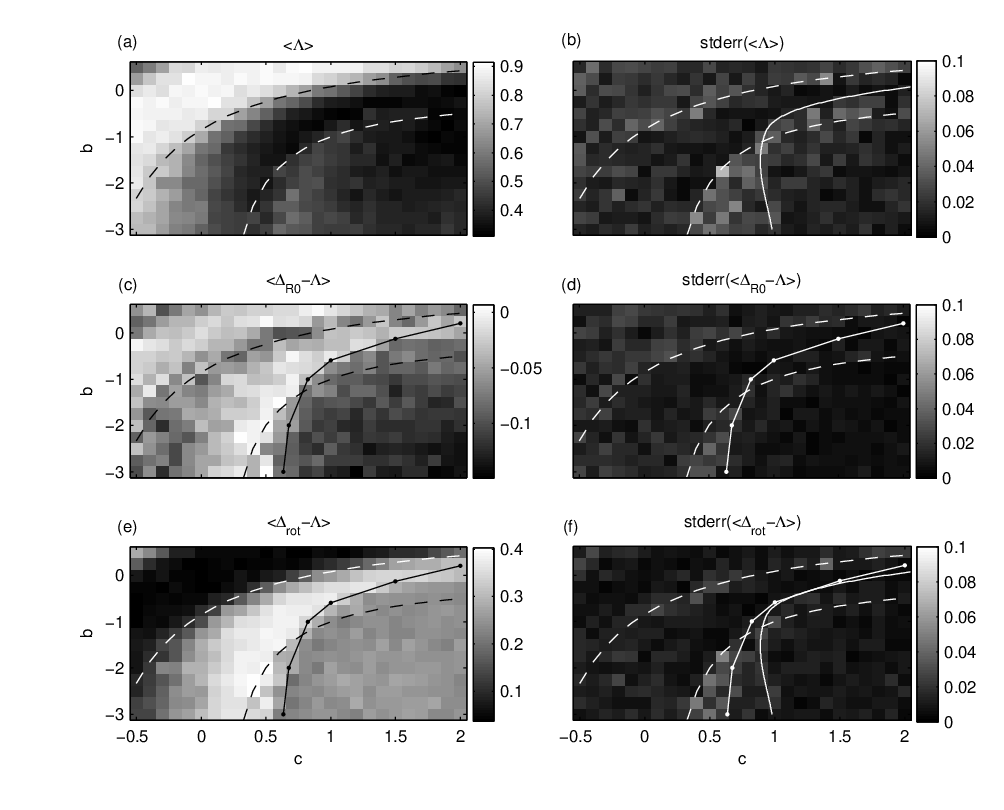}}
\caption{(a) Mean value of the laminarity, $\Lambda$, (b) standard error of the means in a, (c) mean value of the difference of $\Delta_{R0}$ (Fig.~\ref{fig6}c) and $\Lambda$, (d) standard error of the means in c, (e) mean value of the difference of $\Delta_{rot}$ (Fig.~\ref{fig6}e) and $\Lambda$, (f) standard error of the means in e. For each parameter combination there are five independent runs where the real part of the last time stamp is analyzed. The error is given by the ratio of the standard deviation and the square root of five. The lines display transition lines of Fig.~\ref{fig6} a which are reflected by the most clear gradients. The colors are chosen for the highest contrast.}
\label{fig7} 
\end{figure*}

The pictures illustrate typical spatial patterns which reflect different states of the system.
In Fig.~\ref{fig5}a, the plane is dominated by large monotonic areas with a small number of defects, i.e. vortices.
This pattern relates to unbounded states which can be found in the region of $b=c$ of the parameter plane.
Crossing the eponymous line OR from that region (Fig.~\ref{fig6}a), we reach an oscillatory regime where spiral-like structures appear and the distances between their cores, the defects, are in a stable equilibrium which is also called bounded state. 
Here we find the spatial pattern shown in Fig.~\ref{fig5}b-g, for example. 
A lasted change of the parameters $b$ and $c$ toward the line of the Eckhaus instability (EI, c.f. Fig.~\ref{fig6}a) leads to more and more distinctive spirals (Fig.~\ref{fig5}c). 
In the case of random initial conditions, this development ends at line T (c.f. Fig.~\ref{fig6}a) in frozen states, also called vortex glass (Fig.~\ref{fig5}d), where quasi-stationary patterns are formed. 
This property results from a convective transport of perturbations away from the spirals' cores which is theoretical given between the lines of EI and the absolute instability (AI, c.f. Fig.~\ref{fig6}a). 
Beyond the line T, defects are persistently created and annihilated and they less and less emit spiral waves for the change of $b$ and $c$ to the right upper corner of Fig.~\ref{fig6}a. 
Representative spatial pattern of this defect turbulence are given in Fig.~\ref{fig5}e and f, for example. 
Beside this defect turbulence, you can find phase turbulence in the region of the Benjamin-Feir-Newell line (BF, c.f. Fig.~\ref{fig6}a) where no defects occur (Fig.~\ref{fig5}g). 
The range of this region strongly depends on the size of the considered data field.
For a more detailed description of the mentioned states we refer to Aranson \& Kramer~\cite{Aranson2002} and Chat\'e \& Manneville~\cite{Chate1996}.
However, applying the GRP analysis to these patterns we expect that the unbounded states are characterized by a high laminarity and a determinism near one,
whereas spiral-like structures and turbulence patterns should have a small value of $\Lambda$ and a medium value of $\Delta$.   
Taking the random initial conditions into account, we simulate five runs for each considered combination of the parameters ($c={-0.5,-0.4,-0.3,...,2}$ and $b={-3,-2.75,-2.5,...,0.5}$).
For the real part of the last time stamps of each run, we construct the GRP from the central square of each snapshot with the size of $64$ sample points in order to avoid boundary effects of the simulation.
The threshold is fixed to $\epsilon=0.1$ to get recurrence rates in the range from 0.1 to 0.2 as in the regular examples (compare Fig.~\ref{fig6}b with Tab.~\ref{tab2}). 
We calculate $h$, $\Delta_{R0}$, $\Delta_{rot}$ and $\Lambda$ with $l_{min}=2$, and $l_{max}=10$ for the boundary correction.
Finally, the expected values and their standard errors (standard deviation divided by the square root of five) of these measures are estimated over the five runs and plotted in Figs.~\ref{fig6} and \ref{fig7}.

\begin{figure}
\centering
\resizebox{0.4\textwidth}{!}{\includegraphics{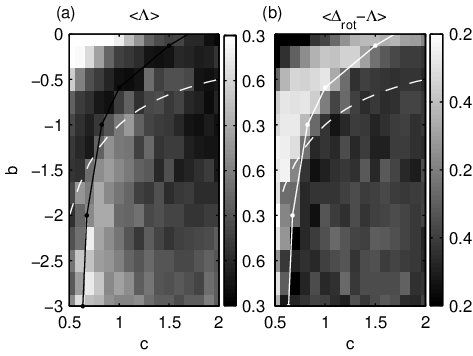}}
\caption{ Zoom of the region of turbulence in Fig.~\ref{fig7}a and e, respectively.}
\label{fig8}
\end{figure}

We see, that the gradients of the traditional measure $\Delta_{R0}$ indicate the line OR, and the region which is enclosed by AI and BF (Fig.~\ref{fig6}c). 
These lines also bound the areas where the standard error of this measure is increased.
In comparison with that, the value of the extended quantifier $\Delta_{rot}$ is clearly larger than $\Delta_{R0}$ in the region below OR (Fig.~\ref{fig6}e).
Here the gradients indicate the transition lines OR and AI, where AI diverges for decreasing values of $b$.
Further the standard error of $\Delta_{rot}$ is remarkable smaller than the standard error of $\Delta_{R0}$ despite the higher mean values (Fig.~\ref{fig6}f).
The comparison of $\Lambda$ and $\Delta_{R0}$ (Fig.~\ref{fig7}a and \ref{fig6}c , respectively) indicates that $\Delta_{R0}$ is strongly affected by the monotonic areas (Fig.~\ref{fig5}) which are quantified by $\Lambda$.
If we assume that the measure of determinism consists of the quantification of monotonic areas in the snapshots and recurrent structured parts then the subtraction of $\Lambda$ from the measure of determinism should uncover the contribution of the latter.
But the negative values of $\Delta_{R0}-\Lambda$ in Fig.~\ref{fig7}c show that the partly consideration of the recurrent structures by $\Delta_{R0}$ leads also to an underestimation of the contribution of the laminar areas to the measure of determinism.
In contrast to that, the use of $\Delta_{rot}$ exhibits the assumed behavior where the difference $\Delta_{rot}-\Lambda$ reveals the expected contributions of the structured patterns (Fig.~\ref{fig5}c-g) to the determinism (Fig.~\ref{fig7}e). 
So, the structured patterns (Fig.~\ref{fig5}c-g) are characterized by a medium value of $\Delta_{rot}-\Lambda$ which is greater than in the case of noise (see Tab.~\ref{tab2}) but remarkable smaller than one.
Further, the higher ordered spiral structures leads to higher values of $\Delta_{rot}-\Lambda$ ($\approx 0.4$) than the turbulent patterns ($\approx 0.2$ to $0.3$)(Fig.~\ref{fig7}e).
We see that the lines OR, the part of BF below the crossing with T and the part of T above this crossing delineate areas of specific levels of determinism.
So, we find the highest values of $\Delta_{rot}-\Lambda$ in the region between OR and the combination of BF, and T which is associated with dominant spiral patterns.
The progression from values up to $0.2$ to $0.4$ reflects the increase of this dominance.
Only in the range of $c\approx -0.5$, there is a slight difference where the gradients indicate a steeper decrease of OR.
Another interesting section is enclosed by BF and T in the lower range of the parameter $b$. 
Here, we identify higher fluctuations in the mean values of $\Delta_{rot}-\Lambda$ (Fig.~\ref{fig7}e) and higher errors (Fig.~\ref{fig7}f) which indicate a coexistence of different patterns getting by chance for the five runs.
These errors are comparable to these ones along the rest of the drawn lines (Fig.~\ref{fig7}f) where intermittent behavior is assumed. 
Finally, there are weak gradients in the lower right corner of the parameter space enclosed by T and BF, the section of turbulence, which indicate different patterns (Fig.~\ref{fig7}a and e).
This tendency is highlighted in Fig.~\ref{fig8}. 
Here, a band of lower values of $\Delta_{rot}-\Lambda$ ($\geq 0.25$) goes along the borders and encloses an area of higher values in the lower right corner (Fig.~\ref{fig8}b).
In contrast to this behavior, $\Lambda$ is continuously increasing from higher values of $b$ to lower ones (Fig.~\ref{fig8}a).
This increase reflects a rising scale of the spatial patterns which is indicated by the snapshots in Fig.~\ref{fig5}f and g.

The shown discriminations impair, if noise affects the patterns. 
So, the $\Delta$ and $\Lambda$ decreases, if the recurrence threshold remains unchanged. 
Therefore, we suggest an adaptation of the threshold by means of strategies which are given in Marwan et al.\cite{Marwan2007a} for the classical recurrence plot analysis.


\section{\label{conclusion} Conclusion}

In this work we have proposed an extended quantification of the GRP by additionally taking into account spatial recurrence by means of rotational symmetry which substantially improves the original approach, which only considers recurrence by means of translational symmetry, and therefore completes the description of circular spatial patterns.
This extension is crucial for a consistent description of complex patterns by means of the GRP as shown in the examples of the complex GLE.
In particular, the examples of regular patterns show that a sufficient evaluation of the spatial regularity, i.e. the predictability, needs the consideration of rotational symmetries in addition to the simple translation (Fig.~\ref{fig3}).
These symmetries result in additional structures formed by 1's in the binary matrix of the GRP causing the found underestimation of $\Delta\def\Delta_{R0}$ which is the base of the diagonal structures.
A combined consideration of these structural elements leads to values of the determinism of $\Delta_{rot}\approx 1$, which are consistent with the theoretical definition.
But this combination has to take into account the overlapping of the structures of 1's in order to avoid multiple counting (Tab.~\ref{tab2}).
The systematic analysis of the more complicated spatial patterns of the complex GLE shows that the extended quantity $\Delta_{rot}$ is more stable than the classical measure $\Delta$ (compare Fig.~\ref{fig6}d and f).
Further, it demonstrates the improved description of circular patterns by means of clearly larger values of $\Delta_{rot}$ in the parameter range of spiral waves and turbulence (below line OR in Fig.~\ref{fig6}c and e).
Therefore, $\Delta_{rot}$ allows a distinction between contributions of constant parts in the data field to the determinism, approximated by $\Lambda$, and actual patterns, quantified by $\Delta_{rot}-\Lambda$, in contrast to $\Delta$ (compare Fig.~\ref{fig7}c and e).
That is, there is a clear separation of the three main regimes: unbounded states, turbulence, and spiral waves; and the expected increase of the determinism of the related actual patterns, from the first to the third regime.

There are, however several important open points for future research, in particular the analysis of the influences of the method's parameters.
Further, we have to investigate the consequences for other measure, which are based on the diagonal structures in the GRP, such as the average size of the diagonal elements.
Finally, the calculating algorithm of $\Delta_{rot}$ needs more optimization than the one of $\Delta$ in order to improve its computational performance.

However, it is worth to remind that all these distinctions shown in the last section are only based on single snapshots of the real part of the Ginzburg-Landau-equation.
This circumstance underlines the power of the proposed method and promises an improved description of dynamical systems originated from fluid dynamics, climatology, biology, ecology, and social sciences.


\begin{acknowledgement}
This work was supported by the Volkswagen Foundation (Grant No. 88462), the DFG RTG 2043/1 "Natural Hazards and Risks in a Changing World".
\end{acknowledgement}


\appendix

\section{\label{allD} General definition of diagonal like structures of GRP}

The general formulation of the group of recurrence patterns with the classical diagonal elements as special case (Eq.~\ref{eq6}) are defined by:
\begin{equation}\label{A4}
\theta\left(\sum_{a=1}^D B_a\right)\prod_{k_1,k_2=b_1}^{b_2}{R_{(\vec{i}+\vec{k},\vec{j}+\vec{k}')}}\equiv 1
\end{equation}
where $b_1$ and $b_2$ are defined by Eq.~\ref{eq10} and Eq.~\ref{eq11}, respectively, and $\vec{i}$, $\vec{k}$, and $\vec{j}$ are $D$ tuples of whole numbers.
The boundary conditions are
\begin{eqnarray}\label{A5}
B_a=&&\sum_{s_1,...,s_{D-1}=b_1-1}^{b_2+1}{(1-R_{(\vec{i}+\vec{u},\vec{j}+\vec{u}')})}\times \nonumber\\
&&\sum_{s_1,...,s_{D-1}=b_1-1}^{b_2+1}{(1-R_{(\vec{i}+\vec{v},\vec{j}+\vec{v}')})}
\end{eqnarray}
where $a$ goes from 1 to $D$.
$\vec{u}$ and $\vec{v}$ are $D$ tuples of whold numbers, too, where the $a$-th component is fixed at $b_1-1$ and $b_2+1$, respectively.
Their components before the $a$-th one, $s_1$ to $s_{a-1}$ , run from  $b_1$ to $b_2$ whereas the components after that run from $b_1-1$ to $b_2+1$.
The tuples $\vec{k}'$, $\vec{u}'$, and $\vec{v}'$ are transformation of $\vec{k}$, $\vec{u}$, and $\vec{v}$, respectively, and correspond to the considered mapping which preserve the relative positions of the grid points among themselves in the $D$ dimensional cuboid.
There are \begin{equation}\label{A6}
\prod_{i=0}^{D-1}{2(D-i)}
\end{equation}
of such operations.
The related transformations we get by combining the components of the original tuple or their negative versions (in the case of odd values of $l$) in a new tuples of the same length where each component or its negation may appears only once.
For even values of $l$ the negative components of the new tuple are extended by adding one.


\section{\label{LAM} Quantification of the vertical structures of GRP}

The vertical structures of 1 in the GRP are quantified by the laminarity 
\begin{equation}\label{eq17}
\Lambda=\frac{\sum_{v=v_{min}}^{N}{v^D P(v)}}{\sum_{v=1}^{N}{v^D P(v)}}
\end{equation}
where $P(v)$ is the histogram of the size $v$ of the $D$ dimensional cuboids. 
$v_{min}$ is equivalent to $l_{min}$ in Eq.~\ref{eq16} and is set to $2$ in our study.


\section{\label{regEx} Regular patterns}

The considered regular patterns are:
\begin{itemize}
	\item Constant values 
	
	The grid is filled with ones.
	\item White noise (e.g. Fig.~\ref{fig3}a) 
	
	The grid is filled with realizations of pairwise independent equal distributed random numbers in the range of $[0,1]$.
	\item Asymmetric regular pattern in one direction\\
	(STRIPE1, Fig.~\ref{fig3}b)\\ 
	$f_{ij}=mod(i,p)$\\
	for all $i=1,\dots,N$ and $j=1,\dots,N$. 
	The period is set to $p=5$ data points.
	\item Asymmetric regular pattern in two directions\\
	(STRIPE2, Fig.~\ref{fig3}c)\\	
	$f_{ij}=mod(i,p_1)+mod(j,p_2)$\\	
	for $i=1,\dots,N$ and $j=1,\dots,N$. 
	The periods in both directions are $p_1=p_2=10$ data points. 
	\item Symmetric regular pattern in one direction\\
	(STRIPE3, Fig.~\ref{fig3}d) \\	
	$f_{ij}=\vert mod(i,p)-2mod(i,p/2)\vert$\\	
	for $i=1,\dots,N$ and $j=1,\dots,N$. 
	The period is $p=10$ data points.
	\item Symmetric regular pattern in two directions\\
	(STRIPE4, Fig.~\ref{fig3}e) \\	
	$f_{ij}=(\vert mod(i,p_1)-2mod(i,p_1/2)\vert\\+\vert mod(j,p_2)-2mod(j,p_2/2)\vert)/2$\\	
	for $i=1,\dots,N$ and $j=1,\dots,N$. The periods in both direction are $p_1=p_2=10$ data points.
	\item Asymmetric regular pattern in one direction reflected on a vertical line (STRIPE5, Fig.~\ref{fig3}f) 
	
	$f_{ij}=mod(mod(\vert i-(N+1)/2\vert,(N+1)/2),p)$
	
	for $i=1,\dots,N$ and $j=1,\dots,N$. The period is $p=5$.
	\item Asymmetric regular pattern in two directions reflected on a vertical line and a horizontal one (STRIPE6, Fig.~\ref{fig3}g) 
	
	$f_{ij}=mod(mod(\vert i-(N+1)/2\vert,(N+1)/2)\\+mod(\vert j-(N+1)/2\vert,(N+1)/2),p)$
	
	for $i=1,\dots,N$ and $j=1,\dots,N$. The periods are $p_1=p_2=p=5$ data points.
	\item Circles (Fig.~\ref{fig3}h) 
	
	$f_{ij}=mod(sqrt((mod(\vert i-(N+1)/2\vert,(N+1)/2))^2\\+(mod(\vert j-(N+1)/2\vert,(N+1)/2))^2),p)$
	
	for $i=1,\dots,N$ and $j=1,\dots,N$. The period is $p=5$ data points.
\end{itemize}


\bibliographystyle{epj}
\bibliography{riedl_2015_circularRP}
%

\end{document}